\documentclass[pra,showpacs,amsmath,amssymb,10pt,twocolumn]{revtex4}

\usepackage{amsthm}
\usepackage{dcolumn}
\usepackage{bm}
\usepackage{graphicx}

\begin{document}

\newcommand\ket[1]{\ensuremath{|#1\rangle}}
\newcommand\bra[1]{\ensuremath{\langle#1|}}
\newcommand\iprod[2]{\ensuremath{\langle#1|#2\rangle}}
\newcommand\oprod[2]{\ensuremath{|#1\rangle\langle#2|}}

\title{Local cloning of two product states}

\author{Zhengfeng Ji}
  \email{jizhengfeng98@mails.tsinghua.edu.cn}
\author{Yuan Feng}
  \email{feng-y@tsinghua.edu.cn}
\author{Mingsheng Ying}
  \email{yingmsh@tsinghua.edu.cn}
\affiliation{
State Key Laboratory of Intelligent Technology and Systems, Department of Computer Science and Technology, Tsinghua University, Beijing 100084, China
}

\date{\today}

\begin{abstract}
Local quantum operations and classical communication (LOCC) put considerable constraints on many quantum information processing tasks such as cloning and discrimination. Surprisingly however, discrimination of any two pure states survives such constraints in some sense. In this paper, we show that cloning is not that lucky; namely, conclusive LOCC cloning of two product states is strictly less efficient than global cloning.
\end{abstract}

\maketitle

\section{Introduction}
Of all the landmark discoveries in quantum computation and quantum information theory, the impossibility of universal cloning~\cite{WZ82,Diek82} has an important position since cloning is one of the most fundamental information processing tasks and the ``non-cloning'' principle has both theoretical and practical inferences. Fruitful results on cloning have also been obtained under the condition when some compromises are made, such as approximate cloning~\cite{BH96,BDE+98} and probabilistic cloning of a finite set of states~\cite{DG98a,DG98b,Pati99,FZY02}. For example, Duan and Guo considered the problem of probabilistic cloning of two non-orthogonal states in Ref.~\cite{DG98a} and later they solved the problem in a more general setting in Ref.~\cite{DG98b}. Since local operations and classical communication (LOCC) were introduced into the research of basic quantum information processing tasks such as cloning and discrimination, more interesting results have been obtained in the literature. These results enrich both the research of quantum information and that of the historical ``non-locality'' problem. Recently, some works on LOCC cloning have been brought up~\cite{GKR04,ACP04,OH04}, all showing that LOCC cloning is somewhat difficult to perform. In this paper, we study the LOCC version of optimal probabilistic cloning and by ``optimal'' we mean that the success probability is maximized.

One other widely studied quantum information processing task which is closely related to cloning is quantum state discrimination. Similar to cloning, perfect discrimination of non-orthogonal states is also impossible and we can only discriminate a finite set of non-orthogonal states probabilistically~\cite{Ivan87,Diek88,Pere88,JS95}. Moreover, discrimination under the constraint of LOCC has also been extensively studied in the literature with even more results obtained than LOCC cloning. For example, it has been shown that there exists a set of orthogonal product states which cannot be discriminated perfectly using LOCC~\cite{BDF+99}. Yet any two pure states, entangled or not, can always be optimally discriminated both conclusively~\cite{WSHV00,CY01,CY02,JCY04} and inconclusively~\cite{VSPM01}. A comparison of these two results leads us to conclude that the inefficiency of LOCC is evident only when there are larger number of unknown states involved in the task.

Keeping in mind the fact that discrimination (or identification) can be regarded as a special case of cloning in which the number of destination copies goes infinite~\cite{DG98b}, we may naturally expect LOCC cloning, as a generalization, to have similar properties of LOCC discrimination. The fact that LOCC can achieve global optimality in discrimination of any two pure states leads us to the question of whether it will remain true in LOCC cloning. Namely, is LOCC still powerful enough to clone pure states efficiently when the number of unknown states is small?

But a simple thought tells that it is almost impossible even if the state to be cloned is exactly known since LOCC cannot increase the entanglement between separated parties. In fact, with Vidal's formula on probabilistic entanglement transformation~\cite{Vida99}, we can see that the success probability of local cloning of a Bell state (or any other entangled pure state) is $0$ while globally we can always clone the state perfectly. One approach that many other authors take to deal with this is to assume that LOCC cloning is performed with the help of a maximally entangled state, or a so-called ``blank state'', acting as a slot where the cloned copy may reside~\cite{LNFJ04,GKR04,ACP04,OH04}. This assumption is natural especially when one of the states to be copied is maximally entangled states. In this paper, however, we focus on a special case of the LOCC cloning problem where only product states are considered. Obviously, the entanglement constraint of LOCC does not apply any more and we can continue to discuss the question raised before nontrivially. Moreover, this separation of LOCC constraints on entanglement is helpful in that otherwise one might wrongly lay the blame on the ``entanglement non-increasing'' property which might be in fact only partly responsible for the inefficiency of LOCC.

As a complete answer to the question, we obtain a formula which precisely calculates how efficient LOCC cloning can be. This formula indicates that it is generally impossible to achieve global optimality in conclusive LOCC cloning even in the simplest case, cloning of two product states with equal prior probability. Namely, for any finite $n$, $m\!\to\!n$ cloning of two nonorthogonal product states cannot be globally optimal using only LOCC; while for infinitely large $n$, the LOCC cloning (which is then a discrimination) can be optimal. From the interesting result proved in Ref.~\cite{BDF+99}, we can see that perfect LOCC cloning of the nine orthogonal product states constructed there is also impossible since otherwise we can clone the state to infinite copies and discriminate them perfectly which has already been proved to be impossible. However, it is easy to see that LOCC can always clone two orthogonal product states perfectly. The inefficiency of LOCC is then further revealed in our paper by analyzing LOCC cloning of two product states when they are non-orthogonal.

Consider two separated parties, Alice and Bob, having $m$ copies of a product state secretly chosen from $\ket{\phi_1}_A \otimes \ket{\phi_2}_B$ and $\ket{\psi_1}_A \otimes \ket{\psi_2}_B$ with equal probability. They want to obtain $n(>m)$ copies of the chosen state, that is, $\ket{\phi_1}^{\otimes n}$ (or $\ket{\psi_1}^{\otimes n}$) on Alice's side and at the same time $\ket{\phi_2}^{\otimes n}$ (resp.\ $\ket{\psi_2}^{\otimes n}$ according to what state they initially possess) on Bob's side. Denote $\mu = |\iprod{\phi_1}{\psi_1}|$, $\nu = |\iprod{\phi_2}{\psi_2}|$ and assume that $\mu < 1$, $\nu < 1$. When Alice and Bob are able to perform arbitrary quantum operations on their joint system, the optimal conclusive cloning~\cite{DG98a,CB98} succeeds with the probability
\begin{equation}\label{eq:global_cloning_prob}
\eta_{c} = \frac{1-\mu^m\nu^m}{1-\mu^n\nu^n}.
\end{equation}
Our main result is that conclusive LOCC cloning of product states cannot achieve this when $\mu \nu$ is not zero which means that the secret states are non-orthogonal. We will revisit the global case problem which is first analyzed in Ref.~\cite{DG98a} and come back to our topic on LOCC cloning later.

Keep the number of initial copies $m$ unchanged and let $n$ tend to infinite. Eq.~\eqref{eq:global_cloning_prob} then gives the optimal success probability of identifying two unknown states when $m$ copies are provided~\cite{Ivan87,Diek88,Pere88,JS95}:
\begin{equation*}
\eta_{d} = 1-\mu^m \nu^m.
\end{equation*}
Moreover, we can have a more general result similar to Eq.~\eqref{eq:global_cloning_prob} in a quantum task called ``quantum state separation'' first introduced in Ref.~\cite{CB98}. Quantum separation generalizes both cloning and discrimination, and thus has a simple and general representation. Suppose we are given $\ket{\phi}$ (or \ket{\psi} as an equal probability alternative), and we want to obtain $\ket{\phi'}$ ($\ket{\psi'}$ respectively) without knowing what state it exactly is. Let $\mu = |\iprod{\phi}{\psi}|$, $\mu' = |\iprod{\phi'}{\psi'}|$ and $\mu \ge \mu'$ which is the key assumption in quantum state separation. It is easy to prove that the maximal probability of success is given by
\begin{equation}\label{eq:global_separation_prob}
\eta_s = \frac{1-\mu}{1-\mu'}.
\end{equation}
To be complete, we prove this formula in the following.

A unitary transformation $U$ on the system and the environment is supposed to be the optimal operation. We expand it in the following:
\begin{subequations}\label{eq:separation}
\begin{eqnarray}
U\ket{\phi}\ket{e} & = & \sqrt{s_1}\ket{\phi'}\ket{e_1} + \sum_{i=2}^n \sqrt{s_i}\ket{\alpha_i}\ket{e_i},\\
U\ket{\psi}\ket{e} & = & \sqrt{t_1}\ket{\psi'}\ket{e_1} + \sum_{i=2}^n \sqrt{t_i}\ket{\beta_i}\ket{e_i},
\end{eqnarray}
where $\ket{e}$ is the initial state of the ancillary system and $\ket{e_i}$ are orthogonal states.
\end{subequations}
Subsequent measurement on the system spanned by $\ket{e_i}$ tells whether the transformation is successful or not. It succeeds with probability $(s_1+t_1)/2$ when outcome is $e_1$ and fails with probability $1 - (s_1+t_1)/2$ otherwise.

The unitary transformation $U$ preserves inner product, that is
\begin{equation}\label{eq:inner_product_preserving}
\iprod{\phi}{\psi} = \sqrt{s_1 t_1} \iprod{\phi'}{\psi'} + \sum_{i=2}^n \sqrt{s_i t_i} \iprod{\alpha_i}{\beta_i}.
\end{equation}
Using the triangle inequality, we have
\begin{eqnarray*}
\mu & \le & \sqrt{s_1 t_1} \mu' + \sum_{i=2}^n \sqrt{s_i t_i} |\iprod{\alpha_i}{\beta_i}|\\
    & \le & \sqrt{s_1 t_1} \mu' + \sum_{i=2}^n \sqrt{s_i t_i}\\
    & \le & \frac{s_1 + t_1}{2} \mu' + \sum_{i=2}^n \frac{s_i + t_i}{2}\\
    & =   & \frac{s_1 + t_1}{2} \mu' + 1 - \frac{s_1 + t_1}{2}.
\end{eqnarray*}
Thus
\begin{equation*}
\eta_s = \frac{s_1 + t_1}{2} \le \frac{1-\mu}{1-\mu'}
\end{equation*}
with equality when $|\iprod{\alpha_i}{\beta_i}| = 1$, $s_i = t_i$ and $\iprod{\phi}{\psi}$, $\iprod{\alpha_i}{\beta_i}$ has the same phase with $\iprod{\phi'}{\psi'}$. The last condition is easily fulfilled as one of the states can be multiplied by a global phase without altering the physical meaning. This completes the proof of Eq.~\eqref{eq:global_separation_prob}.

Then, we study quantum separation of two states with arbitrary prior probability. Generally, it is hard to give an analytical formula for this problem, but we can obtain an upper bound on the success probability. The secret state is now $\ket{\phi}$ with probability $s$ and $\ket{\psi}$ with probability $t$ where $s+t=1$. Again, denote $\mu = |\iprod{\phi}{\psi}|$ and $\mu' = |\iprod{\phi'}{\psi'}|$. Let us consider the expansion in Eq.~\eqref{eq:separation} which also holds though the prior distribution is now not uniform. Thus, we still have
\begin{equation*}
\iprod{\phi}{\psi} = \sqrt{s_1 t_1} \iprod{\phi'}{\psi'} + \sum_{i=2}^n \sqrt{s_i t_i} \iprod{\alpha_i}{\beta_i},
\end{equation*}
and
\begin{equation*}
\mu \le \sqrt{s_1 t_1} \mu' + \sum_{i=2}^n \sqrt{s_i t_i},
\end{equation*}
which gives
\begin{equation*}
\frac{\mu-\mu'}{1-\mu'} \le \frac{(\sqrt{s_1 t_1} - 1) \mu' + \sum_{i=2}^n \sqrt{s_i t_i}}{1-\mu'}.
\end{equation*}
We claim that the right hand side is less than or equal to
\begin{equation*}
\frac{1 - ss_1 - tt_1}{2\sqrt{s t}},
\end{equation*}
for the correctness of which we only need to check
\begin{eqnarray}\label{eq:upper_bound_temp}
& & (2\sqrt{ss_1 tt_1} - 2\sqrt{st} + 1 - ss_1 -tt_1) \mu'\nonumber\\
& \le & 1 - ss_1 - tt_1 - \sum_{i=2}^n 2\sqrt{ss_i tt_i}.
\end{eqnarray}
Notice that the RHS of Eq.~\eqref{eq:upper_bound_temp} is larger than or equal to
\begin{equation*}
1 - \sum_{i=1}^n (ss_i + tt_i) = 0,
\end{equation*}
Eq.~\eqref{eq:upper_bound_temp} holds if the LHS of Eq.~\eqref{eq:upper_bound_temp} is negative. And when the LHS of Eq.~\eqref{eq:upper_bound_temp} is nonnegative, the maximal value of it obtains when $\mu' = 1$, so we need only to prove
\begin{equation*}
2\sqrt{ss_1 tt_1} - 2\sqrt{st} + 1 - ss_1 -tt_1 \le 1 - ss_1 - tt_1 - \sum_{i=2}^n 2\sqrt{ss_i tt_i}
\end{equation*}
which can be further reduced to a Cauchy-Schwartz inequality
\begin{equation}\label{eq:cauchy_inequality}
\sum_{i=1}^n \sqrt{s_i t_i} \le \left( \sum_{i=1}^n s_i \sum_{i=1}^n t_i \right)^{1/2} = 1.
\end{equation}
Thus we have proved that
\begin{equation*}
\frac{\mu-\mu'}{1-\mu'} \le \frac{1 - ss_1 - tt_1}{2\sqrt{s t}},
\end{equation*}
from which our upper bound follows
\begin{eqnarray}\label{eq:separation_upper_bound}
\eta_s = s s_1 + t t_1 \le 1 - 2\sqrt{st}\frac{\mu - \mu'}{1 - \mu'}.
\end{eqnarray}

Having in hand the above results concerning global case of quantum cloning, discrimination and quantum separation, we are now ready to prove our main result by an induction on the number of the maximal possible number of rounds of a protocol. To enjoy the generality which simplifies our proof, we present our proof in the language of quantum separation.

Each one of our players, Alice and Bob, is now restricted to perform arbitrary quantum operation on their own systems but can communicate classically back and forth. Originally, Alice and Bob possess $\ket{\phi_1} \otimes \ket{\phi_2}$ (or $\ket{\psi_1} \otimes \ket{\psi_2}$ with equal probability) and they want to optimally separate it to $\ket{\phi_1'} \otimes \ket{\phi_2'}$ (or $\ket{\psi_1'} \otimes \ket{\psi_2'}$ respectively) using LOCC. Let $\mu = |\iprod{\phi_1}{\psi_1}|$, $\mu' = |\iprod{\phi_1'}{\psi_1'}|$, $\nu = |\iprod{\phi_2}{\psi_2}|$, $\nu'= |\iprod{\phi_2'}{\psi_2'}|$ and $1 > \mu \ge \mu'$, $1 > \nu \ge \nu'$. We will prove that LOCC separation cannot achieve the global separation efficiency
\begin{equation}
\eta_s = \frac{1 - \mu \nu}{1 - \mu' \nu'}
\end{equation}
when $\mu' \nu' \neq 0$ and at least one of $\mu \ge \mu'$ and $\nu \ge \nu'$ is rigorous. In fact, it is first proved that for any LOCC protocol $\mathcal{P}$,
\begin{equation}\label{eq:local_separation_upper_bound}
\eta_s^{\mathcal{P}} \le 1 - \mu \nu + \frac{(1 - \mu)(1 - \nu)}{(1 - \mu')(1 - \nu')} \mu' \nu',
\end{equation}
which is easily verified to be smaller than the global separation efficiency. On the other hand, we will construct a LOCC protocol that achieves the efficiency defined as the RHS of Eq.~\eqref{eq:local_separation_upper_bound}. Combine these two parts we obtain the formula for LOCC separation
\begin{equation}\label{eq:local_separation_efficiency}
\eta_s^{L} = 1 - \mu \nu + \frac{(1 - \mu)(1 - \nu)}{(1 - \mu')(1 - \nu')} \mu' \nu'.
\end{equation}

In order to prove the upper bound in Eq.~\eqref{eq:local_separation_upper_bound}, we prove a more general bound that allows unequal initial distribution. Let $s$ and $t$ be the initial distribution and $\eta_s^{\mathcal{P}}(s,t)$ be the efficiency of LOCC separation protocal $\mathcal{P}$. We will show
\begin{equation}\label{eq:general_uppper_bound}
\eta_s^{\mathcal{P}}(s,t) \le 1 - 2\sqrt{st} \mu\nu + 2\sqrt{st} \frac{(1-\mu)(1-\nu)} {(1-\mu')(1-\nu')} \mu' \nu',
\end{equation}
for any protocoal $\mathcal{P}$ and any $s$ and $t$. It is easily seen that when $s=t=1/2$, Eq.~\eqref{eq:general_uppper_bound} degenerates to Eq.~\eqref{eq:local_separation_upper_bound}.

Now, consider any LOCC protocol $\mathcal{P}$ that separates $\ket{\phi_1} \otimes \ket{\phi_2}$ with probability $s$ (or $\ket{\psi_1} \otimes \ket{\psi_2}$ with probability $t$) to $\ket{\phi_1'} \otimes \ket{\phi_2'}$ (or $\ket{\psi_1'} \otimes \ket{\psi_2'}$). In such a LOCC protocol, local operations and classical communication can be carried out repeatedly in arbitrarily many rounds. For example, Alice goes first by measuring her part of system and informing the outcome to Bob, then Bob performs a measurement corresponding to the information Alice tells him, and so on. If we define a round to be a measurement on one's side together with a notification of the result to the other side, any execution (a concrete experiment) of the LOCC protocol is just a sequence of many rounds. Different executions of a same protocol may consist of different number of rounds since each round, except the first one, depends highly on the outcomes of the previous rounds. We prove the upper bound in Eq.~\eqref{eq:general_uppper_bound} by induction on the maximal possible number of rounds of any LOCC protocol $\mathcal{P}$.

To see that Eq.~\eqref{eq:general_uppper_bound} holds when $\mathcal{P}$ contains at most one round, we will employ the upper bound on global separation in Eq.~\eqref{eq:separation_upper_bound} proved before. Without loss of generality, let Alice perform the only round in protocol $\mathcal{P}$. What Bob can do is then merely some unitary transformation on his system which preserves the inner product $\nu$. Thus, when $\nu$ and $\nu'$ are not equal, $\mathcal{P}$ fails definitely and Eq.~\eqref{eq:general_uppper_bound} is obvious. When $\nu = \nu'$, protocol $\mathcal{P}$ succeeds if and only if Alice successfully performs separation on her side. Eq.~\eqref{eq:separation_upper_bound} says that Alice's chance to make it is bounded by
\begin{equation*}
1 - 2\sqrt{st}\frac{\mu - \mu'}{1 - \mu'},
\end{equation*}
which is exactly what we want when substituting $\nu = \nu'$ into Eq.~\eqref{eq:general_uppper_bound}. This establishes the initial condition of our inductive proof.

Each protocol $\mathcal{P}$ with maximally $l$ rounds can be reduced to protocols that have at most $l-1$ rounds after the first round being performed by one of our players, say, Alice. Let $\{ M_i \}$ be the measurement carried out by Alice in the first round of  protocol $\mathcal{P}$; let $s_i$ and $\ket{\phi_1^i}$ be the probability and the post-measurement state respectively when the observed result is $i$ and the secret state of her system is actually $\ket{\phi_1}$; let $t_i$ and $\ket{\psi_1^i}$ be the correspondences when her part of secret state is prepared in $\ket{\psi_1}$. That is,
\begin{subequations}\label{eq:measurement}
\begin{eqnarray}
M_i \ket{\phi_1} & = & \sqrt{s_i} \ket{\phi_1^i}\\
M_i \ket{\psi_1} & = & \sqrt{t_i} \ket{\psi_1^i},
\end{eqnarray}
where $\sum_i M_i^{\dagger} M_i = I$.
\end{subequations}
Let $p_i$ be the probability that result $i$ occurs, $p_{\phi | i}$ and $p_{\psi | i}$ be the new distributions of the secret state, then
\begin{eqnarray*}
p_i & = & s s_i + t t_i\\
p_{\phi | i} & = & \frac{s s_i}{p_i}\\
p_{\psi | i} & = & \frac{t t_i}{p_i}.
\end{eqnarray*}
Efficiency of protocol $\mathcal{P}$ then equals to
\begin{equation*}
\eta_s^{\mathcal{P}} = \sum_i p_i \eta_s^{\mathcal{P}^i}(p_{\phi | i}, p_{\psi | i}),
\end{equation*}
where $\eta_s^{\mathcal{P}^i}(p_{\phi | i}, p_{\psi | i})$ is the efficiency of the further separation $\mathcal{P}^i$ when result $i$ occurs in the above measurement. Since $\mathcal{P}^i$ has at most $l-1$ rounds, by induction hypothesis, we can continue the last equation with
\begin{eqnarray*}
\eta_s^{\mathcal{P}} & = & \sum_i p_i \eta_s^{\mathcal{P}^i}(p_{\phi | i}, p_{\phi | i})\\
                     & \le & \sum_i p_i \left[1 - 2\sqrt{p_{\phi | i} p_{\phi | i}} \left( \mu_i \nu - \frac{(1-\mu_i)(1-\nu)}{(1-\mu')(1-\nu')} \mu' \nu' \right) \right]\\
                     & = & 1 - 2\sqrt{st} \left(\sum_i \sqrt{s_i t_i} \mu_i \right) \nu +\\
                     &   & 2\sqrt{st} \left(\sum_i \sqrt{s_i t_i} - \sum_i \sqrt{s_i t_i} \mu_i \right) \frac{(1-\nu)\mu'\nu'}{(1-\mu')(1-\nu')},
\end{eqnarray*}
where $\mu_i = |\iprod{\phi_1^i}{\psi_1^i}|$.
Employing Eq.~\eqref{eq:measurement}, we have
\begin{equation*}
\mu = |\sum_i \sqrt{s_i t_i} \iprod{\phi_1^i}{\psi_1^i}| \le \sum_i \sqrt{s_i t_i} \mu_i.
\end{equation*}
From the above equation and Eq.~\eqref{eq:cauchy_inequality}, we obtain
\begin{equation*}
\eta_s^{\mathcal{P}} \le 1 - 2\sqrt{st} \mu \nu + 2\sqrt{st} \frac{(1-\mu) (1-\nu)}{(1-\mu') (1-\nu')} \mu' \nu',
\end{equation*}
which completes the proof.

Returning back to the case when the initial distribution is uniform (that is, $s=t$), we have the upper bound
\begin{equation*}
\eta_s^{\mathcal{P}} \le 1 - \mu \nu + \frac{(1-\mu) (1-\nu)}{(1-\mu') (1-\nu')} \mu' \nu',
\end{equation*}
by substituting $s$ and $t$ for $1/2$. The interesting thing is that such an upper bound is also achievable. We construct a protocol $\mathcal{P'}$ to show this. In protocol $\mathcal{P'}$, Alice and Bob optimally separate their own part first. If both of them succeed, the protocol finishes, else if only one of them succeeds, he (she) then performs optimal discrimination of the separated state and tells the result to the other one if the discrimination is again successful. Otherwise, the procedure fails. Then the probability of success is
\begin{eqnarray*}
\eta_s^{\mathcal{P'}} & = & \frac{(1-\mu)(1-\nu)}{(1-\mu')(1-\nu')} + \left(1 - \frac{1-\mu}{1-\mu'} \right) \frac{1-\nu}{1-\nu'} (1-\nu') +\\
                      &   & \frac{1-\mu}{1-\mu'} \left(1 - \frac{1-\nu}{1-\nu'} \right) (1-\mu')\\
                      & = & 1 - \mu\nu + \frac{(1-\mu)(1-\nu)}{(1-\mu')(1-\nu')} \mu'\nu',
\end{eqnarray*}
which coincides with our upper bound. Then we get the formula that calculates the efficiency of LOCC separation of two product states:
\begin{equation*}
\eta_s^{L} = 1 - \mu \nu + \frac{(1-\mu) (1-\nu)}{(1-\mu') (1-\nu')} \mu' \nu'.
\end{equation*}
It is easily seen to be strictly less than the global efficiency when $0 < \mu' < \mu < 1$ and $0 < \nu' < \nu < 1$.

Via a simple substitution, we get the efficiency of LOCC cloning
\begin{equation*}
\eta_c^{L} = 1 - \mu^m \nu^m + \frac{(1-\mu^m) (1-\nu^m)}{(1-\mu^n) (1-\nu^n)} \mu^n \nu^n,
\end{equation*}
which is also strictly less than the global efficiency since $0 < \mu^n < \mu^m < 1$ and $0 < \nu^n < \nu^m < 1$ is obvious in an $m\!\to\!n$ cloning. The corresponding optimal protocol $\mathcal{P'}$ becomes that Alice and Bob perform $m\!\to\!n$ cloning separately and if any one of them fails, they resort to the discrimination protocol to improve the efficiency.

In sum, we have analyzed the problem of LOCC cloning and LOCC separation of two product states. It is proved that, except some trivial cases, $m\!\to\!n$ LOCC cloning of product states is less efficient than global cloning. This result strongly contrasts with the fact that any two pure states can be locally discriminated with global efficiency. In other words, LOCC pose more constraints on cloning than on discrimination. This is accordant with some recent results of the related works~\cite{ACP04,OH04}. Since product state does not involve us in the entanglement restriction of LOCC, our result also shows that there is something else, other than the entanglement constraint, that obstructs the cloning procedure in LOCC.

Efficiency, the success probability in conclusive cloning and discrimination, is a natural and important measure. Then why do cloning and discrimination (a cloning with infinite destination copies) have such a difference in it? One observation is that discrimination is somewhat classical since the final result it cares is classical while cloning is not. So in LOCC discrimination of two product states, if either of two parties succeeds then the whole task is done by communicating the result while in LOCC cloning, both of them are required to succeed on their own. Alternatively, from the view of quantum separation, discrimination is a separation where $\mu'$ or $\nu'$ equals to $0$ which makes it a ``trivial'' task while cloning is not trivial generally.

We are thankful to the colleagues in the Quantum Computation and Quantum Information Research Group for helpful discussions. This work was supported by the Natural Science Foundation of China (Grant Nos. 60273003, 60433050, and 60305005).

\bibliography{locop}

\end{document}